\documentstyle[twoside,fleqn,espcrc2]{article}

\newcommand{\AmS}{{\protect\the\textfont2
  A\kern-.1667em\lower.5ex\hbox{M}\kern-.125emS}}

\input epsf
\epsfverbosetrue
\title{\vspace*{-5mm}\rightline{\small  HU--TFT--96-28}
  \vspace*{-3mm}
  \rightline{\small August 4, 1996}
  Four Quarks from Lattice to the Continuum}

\author{Petrus Pennanen\address{Research Institute for Theoretical Physics,
        P.O. Box 9, FIN-00014 University of Helsinki, Finland}
	\thanks{e-mail: {\tt Petrus.Pennanen@helsinki.fi}}}

\begin{document}

\begin{abstract}
A continuum extrapolation of static four- and two-quark energies calculated in 
quenched SU(2) is done based on Sommer's method of setting the scale. A model 
for four-quark energies with explicit gluonic degrees of freedom removed is 
fitted to these energies and the behavior of the parameters of the model is 
investigated.
\end{abstract}

\maketitle

\section{\bf Introduction}

Our group is attempting to understand hadronic 
interactions from first principles by developing a potential model for 
multi-quark systems with explicit gluonic degrees of freedom removed 
(\cite{glpm:96} and references therein). Such a 
model would shed light on multi-quark bound states and e.g. meson-meson 
scattering. 

We have been simulating systems of four static quarks in quenched SU(2) at 
$\beta=2.4$ with a $16^3\times 32$ lattice and the Wilson action. Various 
geometries such as rectangular, linear and tetrahedral have been considered to 
get a set representing the general case.
 
To find the lattice artifacts in energies and their parameterization,
I have performed continuum extrapolations based on Sommer's 
method of setting the scale \cite{som:94} at $\beta=2.35, 2.4$ 
($16^3\times 32$), 2.45 ($20^3\times 32$), 2.5 ($24^3\times 32$) and 
2.55 ($26^3 \times 32$) for squares and tilted rectangles, since these 
geometries exhibit the largest binding energies.

\section{A Model for Four-Quark Energies}

The model is based on static two-quark potentials for different pairings of 
the four-quarks. In the case of two pairings or basis states A,B, the 
eigenenergies $\lambda_i$ are obtained
by diagonalising
\begin{equation}
\label{Ham}
\left[{\bf V}-\lambda_i {\bf N}\right]\Psi_i=0, \label{emodel}
\end{equation}
with
\begin{eqnarray}
\label{NV}
{\bf N} & = & \left(\begin{array}{ll}
1&f/N_c\\
 f/N_c&1\end{array}\right)\ \ {\rm and} \\ \nonumber
{\bf V} & = & \left(\begin{array}{cc}
v_{13}+v_{24} & \frac{f}{N_c}V_{AB}\\
\frac{f}{N_c}V_{BA}&v_{14}+v_{23}\end{array}\right), \label{env}
\end{eqnarray}
where $v_{ij}$ represents the static two-body potential between quarks $i$ and 
$j$ and $V_{AB}$ is from the perturbative expression
$V_{ij}=-{\cal N}(N_c) {\bf T_i} \cdot {\bf T_j} v_{ij}$. The normalization 
is chosen to give for a color singlet state $[ij]^0$ 
$<[ij]^0|V_{ij}|[ij]^0>=v_{ij}$. The four-quark binding energies $E_i$ are 
obtained by subtracting 
the internal energy of the basis state with the lowest energy, e.g. 
$
E_i = \lambda_i-(v_{13}+v_{24}).
$

A central 
element in the model is the phenomenological factor $f$ appearing in the 
overlap of the basis states $<A|B> = f/N_c$ for $SU(N_c)$. 
\begin{itemize}
\item $f$ is a function of the coordinates of all quarks, making the 
$\frac{f}{N_c}V_{AB}$ in eq. \ref{env} a \underline{four-body potential}
\item in the weak coupling limit $f=1$
\item a parameterization $f=f_c e^{-k_A b_S A-k_P \sqrt{b_S} P}$, where $A,P$ 
are the minimal area and perimeter, $b_S$ the string tension and $f_c,k_A,k_P$
are to be fitted
\end{itemize}
Perturbative calculation to $O(\alpha^2)$ produces the two-state model of eq. 
\ref{emodel} with $f=1$ \cite{lan:95}.

The above model using only the ground state of the two-body potential works 
for rectangles but fails for tetrahedrons.

\section{\bf Setting the Scale}

Sommer \cite{som:94} has designed a popular new way to set the scale that 
avoids the long-distance limit in the definition of the string tension
$b_S = \lim_{r\rightarrow\infty} F(r)$. After calculating the force $F(r/a)$ 
between two static quarks, the equation 
\begin{equation}
(r_0/a)^2 F(r_0/a) = c \label{esommer}
\end{equation}
is solved with $c=1.65$ for $r_0/a$. Equating to the continuum value 
$r_0\approx 0.5$ fm gives $a$. 

Choosing $c=1.65$ produces agreement on $r_0=0.49$ fm for 
the Cornell \cite{eic:80} and Richardson \cite{ric:79} nonrelativistic 
effective potentials, while other 
models \cite{mar:81,lee:90} give $r_0$ from $0.44$ to $0.56$ fm. However,
setting $c=2.44$ makes 
three potential models agree on $r_0=0.66$ fm while others give $0.625$ 
or $0.64$ fm. 

Runs were performed for square and tilted rectangle (TR) 
geometries, with resulting scales shown in table \ref{tri}. Here $a$ is 
calculated using $c=1.65$, \\ $a^{\rm II}$ using $c=2.44$ and $a^{b_S}$ from
the string tension using a continuum value $\sqrt{b_S}=0.44$ GeV. 
We can see that at each $\beta$, $a^{b_S}/a\approx 1.09$, and becomes 
$\approx 1$ if we set $\sqrt{b_S}=0.478(4)$ GeV. Meanwhile $a^{II}$ needs no
such adjustment to agree with $a^{b_S}$. 

So it seems $c=2.44$ is a better choice than $c=1.65$. I chose the 
former value, keeping in mind that the difference can be accidental
due to the significant uncertainties in the nonrelativistic potential 
models and the difference of the quenched SU(2) string tension from a
phenomenological value.

\begin{table}[htb]
\begin{center}
\begin{tabular}{lccc} \hline
$\beta$  & $a$ [fm]   & $a^{\rm II}$ [fm] & $a^{b_S}$ [fm] \\ \hline
2.35     & 0.1302(6)  & 0.1401(14) & 0.1408(11)  \\
2.35(TR) & 0.1299(7)  & 0.1397(6)  & 0.1402(5)  \\
2.4      & 0.1101(10) & 0.1186(10) & 0.1194(9) \\
2.4(TR)  & 0.1103(5)  & 0.1190(4)  & 0.1200(11)  \\
2.45     & 0.0918(7)  & 0.0997(8)  & 0.0997(7)  \\
2.45(TR) & 0.0929(7)  & 0.1002(7)  & 0.1010(6)  \\
2.5      & 0.0793(5)  & 0.0856(9)  & 0.0866(9)  \\ 
2.5(TR)  & 0.0789(5)  & 0.0851(6)  & 0.0859(6)  \\ 
2.5(W)   & 0.0793(3)  & 0.0856(6)  & 0.0864(5) \\
2.55     & 0.0675(5)  & 0.0730(5)  & 0.0741(7)  \\
2.55(TR) & 0.0673(9)  & 0.0727(10) & 0.0734(8)  \\ \hline
\end{tabular}
\caption{Values of $a$ for each $\beta$. (W) is for Wuppertal. \label {tri}}
\end{center}
\end{table}

\section{Extrapolating Two-Body Potentials...}

\begin{figure}[htb] 
\epsfxsize=200pt\epsfbox{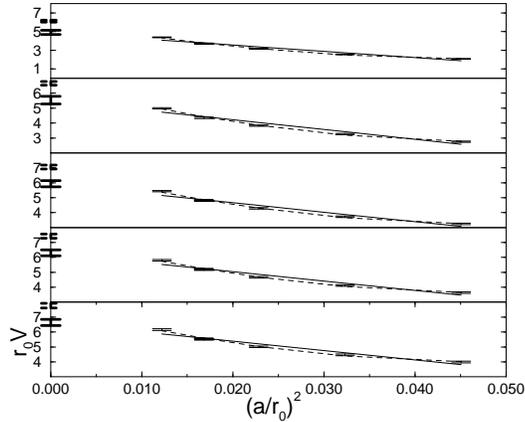}
\caption{Two-body potentials for squares \label{ftrv} with length from 
$0.17$ (top) to $0.52$ fm (bottom)}
\end{figure}

To extrapolate, we need values of energies at different $\beta$'s 
corresponding to the same physical size.  
Sample two-body extrapolations (from a total of 29 potentials) are shown in 
fig. \ref{ftrv} for potentials with separation from 0.17 to 0.52 fm 
involved in squares. Other potentials look similar. A quadratic fit is 
clearly preferred.

\section{...and Four-Body Binding Energies} 

\begin{figure}[htb]
\hspace{0.3cm}\epsfxsize=200pt\epsfbox{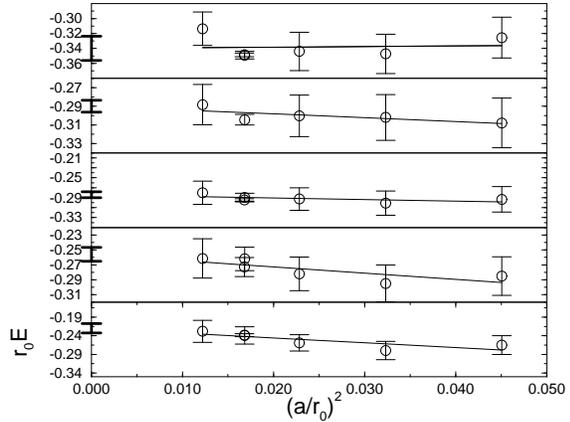}
\caption{Four-body binding energies for squares \label{fsqu}}
\end{figure}

Binding energies for the simulated geometries can be interpolated using the 
$f$-model. I used the parameterization $f=f_c e^{-k_A b_S A}$, as the perimeter
term was not needed. Results for squares with the length of a side 
from 0.17 to 0.52 fm are shown in fig. \ref{fsqu}. Results for tilted 
rectangles look similar. The weaker dependence on $(r_0/a)^2$ shows that there 
are significantly less lattice artifacts than for the two-body potentials.

\section{ Extrapolating the Parameters of the Model}

\begin{figure}[hbt]
\hspace{0cm}\epsfxsize=200pt\epsfbox{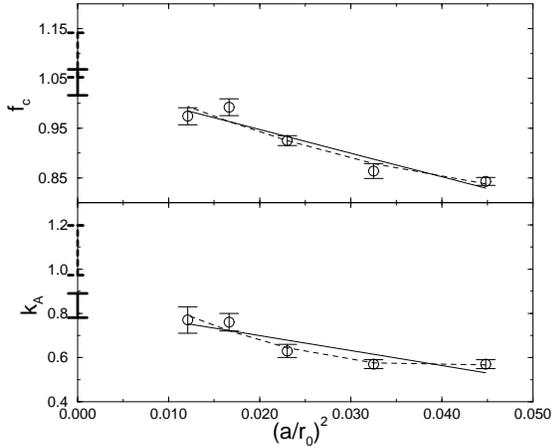}
\caption{$f_c$ and $k_A$ continuum extrapolations}
\label{fsomfc}
\end{figure}

Values of the parameters were obtained by fitting $f=f_c e^{-k_A b_S A}$ 
to the energies of squares and tilted rectangles together at each $\beta$. 
The normalization $f_c$ and the 
constant multiplying the area, $k_A$, are extrapolated in fig. \ref{fsomfc}.

The parameters are changing quite a lot, while $f$ itself is more constant at 
different $\beta$'s. This can be taken as a sign of a poor parameterization. 

\section{ Continuum Fit of the Model}

After extrapolating the two- and four-body energies the $f$-model was fitted 
to the continuum values. Both linearly and quadratically extrapolated 
energies were fitted. Results of the most relevant fits and parameter 
extrapolations are shown in table \ref{tcfit}.

\begin{table}[htb]
\begin{center}
\begin{tabular}{lccc} \hline
type                      & $f_c$    & $k_A$      & $\chi^2$ \\ \hline
lin/quad          & 1.04(4)  & 1.04(6)    & 0.7 \\
lin/quad, $f_c=1$ & 1        & 0.98(3)    & 0.8 \\
quadratic             & 1.08(8)  & 1.37(12)   & 0.4  \\
quadratic, $f_c=1$    & 1        & 1.26(6)    & 0.5  \\
linear extrap.          & 1.04(3)  & 0.83(6)    & 2.6/3.3 \\
quadratic extrap.       & 1.10(5)  & 1.09(11)   & 2.7/0.4 \\  \hline 
\end{tabular}
\caption{Continuum fit and parameter extrapolation results. Fit type 
``lin/quad'' denotes linear extrapolations for four-body energies and 
quadratic for two-body potentials.\label{tcfit}}
\end{center}
\end{table}

Our best estimate for $k_A$ is slightly above one, while $f_c$ can be safely 
set to one. 

In the strong coupling limit
a factor $\exp{(-b_S A)}$, similar to $f$, appears in the off-diagonal 
elements of the Wilson loop matrix, $A$ being the minimal surface bounded by 
straight
lines connecting the quarks. It has been argued that weaker couplings lead 
to a smaller transition area \cite{mat:87}. If this is the case, it is not 
reflected in our best estimate of $k_A$. The larger transition area pointed 
to by $k_A>1$ 
could be explained by the finite width of flux tubes as compared to the lines
in the strong coupling approximation. 

Continuum fits were also performed for squares extrapolated using the original
value $c=1.65$ in equation \ref{esommer}. The resulting values of 
$f_c$ and $k_A$ for fits to all linear and quadratic data were within errors 
with fits to extrapolations using $c=2.44$.

\section{ How to develop the model?}

The failure of the simplest form of the $f$-model to predict the binding 
energies of the 
tetrahedral geometry has been proposed to be due to a dependence of the 
four-body
ground state energy $E_4$ on the {\bf excited state of the two-body potential}.
This was investigated using three fuzzing levels, all with $c=4$, producing 
different ground- and excited state overlaps of the two-body paths. 
\begin{itemize}
\item fuzz level 150: overlaps 97.9-98.7 \%, intermediate convergence of $E_4$
\item fuzz level 20: overlaps 99.1-100 \%, best convergence of $E_4$
\item fuzz level 0: overlaps 85.3-98.2 \%, worst convergence of $E_4$
\end{itemize}
Thus the excited state contribution to $E_4$ comes from the overlap of the
ground state of a gluon field between two quarks with the excited state of 
a gluon field of another quark pair. In ref. \cite{pen:96b} this is  
investigated further and a generalization of the model is introduced that is 
capable of 
reproducing the degenerate ground state energy of a regular tetrahedron.

It is evident that using a high fuzzing level 150 (as in Wuppertal 
\cite{bal:94}) does not necessarily lead to better ground state overlaps on a 
$24^3$ spatial lattice.

The {\bf effects of instantons} are another candidate for a missing feature 
in our model.
Instanton liquid models include neither perturbative one-gluon exchange
(Coulomb part of the two-quark potential) nor confinement
(linear part of said potential). They still predict the nucleon
to be bound with the right mass \cite{sch:96}. Our model gets the energy scale
only from the two-body potential and the overlap factor $f$ can hardly account
for instanton effects.

Our next project is an attempt to uncover the microscopic origins of the 
simulation results by relating the model to flux distributions via energy
sum-rules similar to those derived by C. Michael \cite{mic:95}.

\section{ Acknowledgments}

I warmly thank A.M. Green for support and encouragement, J. Lukkarinen and 
P. Laurikainen for discussions 
and C. Schlichter from Wuppertal for giving us their simulation 
results. Funding from the Finnish academy is gratefully acknowledged. Our 
simulations 
were performed at the Center for Scientific Computing in Helsinki.

\end{document}